\documentclass[num-refs]{nbdt-article}

\DeclareFloatingEnvironment[name={Supplementary Figure}]{suppfigure}

\papertype{Original Article}
\paperfield{Journal Section}

\title{Explaining the effectiveness of fear extinction through latent-cause inference}

\author[1]{Mingyu Song}
\author[2,3,4]{Carolyn E. Jones}
\author[4]{Marie-H. Monfils}
\author[1,5]{Yael Niv}
\affil[1]{Princeton Neuroscience Institute, Princeton University, Princeton, NJ 08540}
\affil[2]{Department of Neurology, Oregon Health \& Science University, Portland, OR 97239}
\affil[3]{Research \& Development, VA Portland Healthcare System, Portland, OR 97239}
\affil[4]{Department of Psychology and Institute for Mental Health Research, The University of Texas at Austin, Austin, TX 78712}
\affil[5]{Department of Psychology, Princeton University, Princeton, NJ 08540}

\corraddress{Mingyu Song}
\corremail{mingyus@princeton.edu}

\fundinginfo{MS and YN were supported by the National Institute of Drug Abuse (R01DA042065).}

\runningauthor{Song et al.}

\begin{document}

\maketitle

\begin{abstract}
Acquiring fear responses to predictors of aversive outcomes is crucial for survival. At the same time, it is important to be able to modify such associations when they are maladaptive, for instance in treating anxiety and trauma-related disorders. Standard extinction procedures can reduce fear temporarily, but with sufficient delay or with reminders of the aversive experience, fear often returns. The latent-cause inference framework explains the return of fear by presuming that animals learn a rich model of the environment, in which the standard extinction procedure triggers the inference of a new latent cause, preventing the unlearning of the original aversive associations. This computational framework had previously inspired an alternative extinction paradigm -- gradual extinction -- which indeed was shown to be more effective in reducing the return of fear. However, the original framework was not sufficient to explain the pattern of results seen in the experiments. Here, we propose a formal model to explain the effectiveness of gradual extinction in reducing spontaneous recovery and reinstatement effects, in contrast to the ineffectiveness of standard extinction and a gradual reverse control procedure. We demonstrate through quantitative simulation that our model can explain qualitative behavioral differences across different extinction procedures as seen in the empirical study. We verify the necessity of several key assumptions added to the latent-cause framework, which suggest potential general principles of animal learning and provide novel predictions for future experiments.

\end{abstract}

\section{Introduction}

Fear memories are notoriously hard to erase. After an association has been formed between an originally neutral cue (e.g., a tone) and some aversive outcome (e.g., a foot shock), animals cannot simply unlearn this association through standard extinction procedures (i.e., being presented with the cue repeatedly in the absence of the aversive outcome) \cite{dunsmoor2015rethinking}. During extinction, the animals' fear response towards the cue gradually reduces, but it usually returns if tested after a long delay (spontaneous recovery)\cite{pavlov1927conditioned, rescorla2004spontaneous}, or if the animal is reminded of the aversive outcome (reinstatement)\cite{pavlov1927conditioned, rescorla1975reinstatement}. Associative learning theory explains these phenomena by postulating that extinction involves learning a {\em new} association rather than updating the original fear association \cite{bouton2004context}. However, this theory does not delineate the particular circumstances under which a new association is initiated, and how this can be avoided so that the original association may be modified through new experience. 

Recent theoretical work \cite{gershman2010context} suggested a formalization of the way animals decide when to learn a new association and when to update old associations using a \emph{latent-cause inference} framework. In this framework, all observations (e.g., cues and reinforcers) are assumed to be generated by latent (unobservable) causes that are each active for a certain (unknown) amount of time. Each latent cause has a certain tendency to generate observations, characterized by its ``generative strength''\footnote{In associative learning, the tendency for two stimuli to co-occur is characterized by an ``associative strength''. Inspired by this, here we term the tendency for a latent cause to generate observations ``generative strength'' to emphasize the causal/generative relationship. This concept is also similar to the ``emission probability'' in Hidden Markov Models.} of each observation. While interacting with the environment, animals infer what latent cause is currently active based on their prior experience and current observations, and behave accordingly. At the same time, they learn and update their estimates of the generative strength of the currently active latent cause.

According to this latent-cause inference framework, in fear extinction, animals infer that there are two distinct latent causes, based on their distinct tendency to generate shocks: one dangerous latent cause (active during conditioning, with a high probability of generating shocks), and one safe latent cause (active during extinction, with low or no probability of generating shocks). The reduction of fear response during extinction is a result of the animal's increasing belief that the second cause is active as more no-shock observations accumulate. Presentation of a shock in a reinstatement procedure is taken to indicate that the original dangerous cause is likely to be active again, causing the return of fear. Similarly, after some passage of time, both causes (dangerous and safe) are equally likely to be active again, leading to the spontaneous recovery of fear response as compared to that measured shortly after extinction. 

\begin{figure}[h!]
\centering
\includegraphics[width=0.8\textwidth]{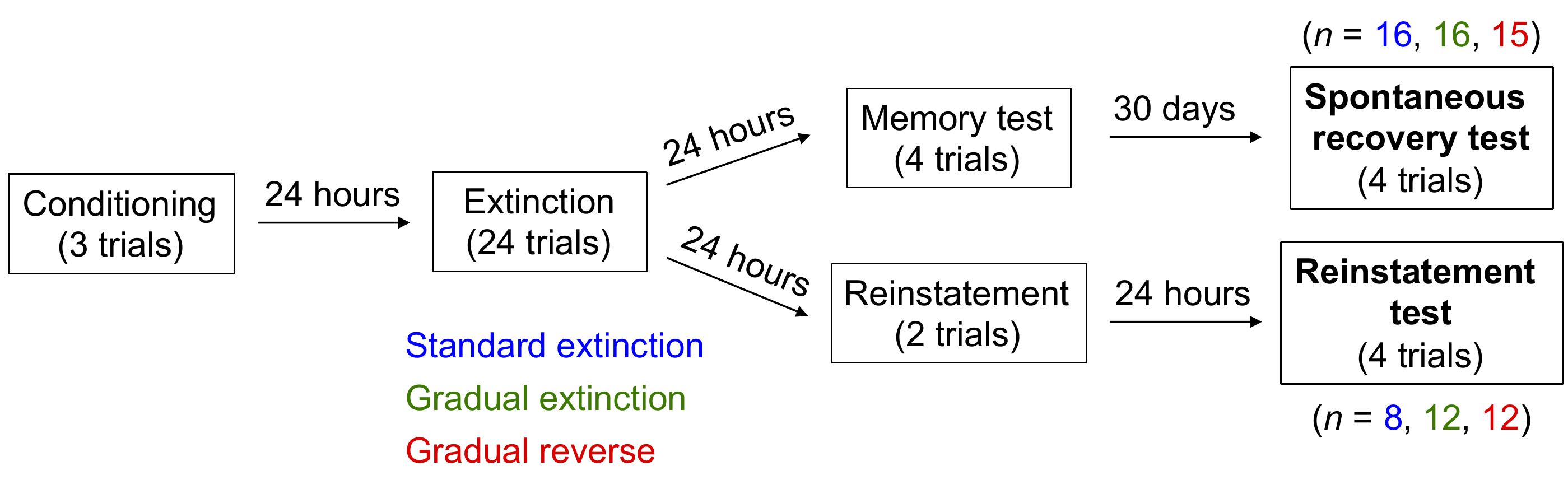}
\caption{\textbf{Experimental design in Gershman et al. \cite{gershman2013gradual}.} Across two experiments (spontaneous recovery and reinstatement), rats were assigned to three different extinction conditions: standard extinction, gradual extinction and gradual reverse. All animals first underwent a conditioning session (3 trials of tone-shock pairing). This was followed by an extinction session (procedures differed based on the extinction condition) 24 hours later. In the spontaneous recovery experiment, animals were first tested on their memory of extinction after another 24 hours (termed a ``long-term memory test'' in the original paper), and then tested for spontaneous recovery of fear response 30 days later. In the reinstatement experiment, animals underwent 2 reinstatement trials (shocks presented alone) 24 hours after extinction, and then were tested for fear response after another 24 hours. The number of animals participating in each experimental condition is noted for each experiment, color-coded based on the extinction procedure: standard extinction in blue, gradual extinction in green, and gradual reverse in red.}
\label{fig:exp_design}
\end{figure}

\begin{figure}[h!]
\centering
\includegraphics[width=0.7\textwidth]{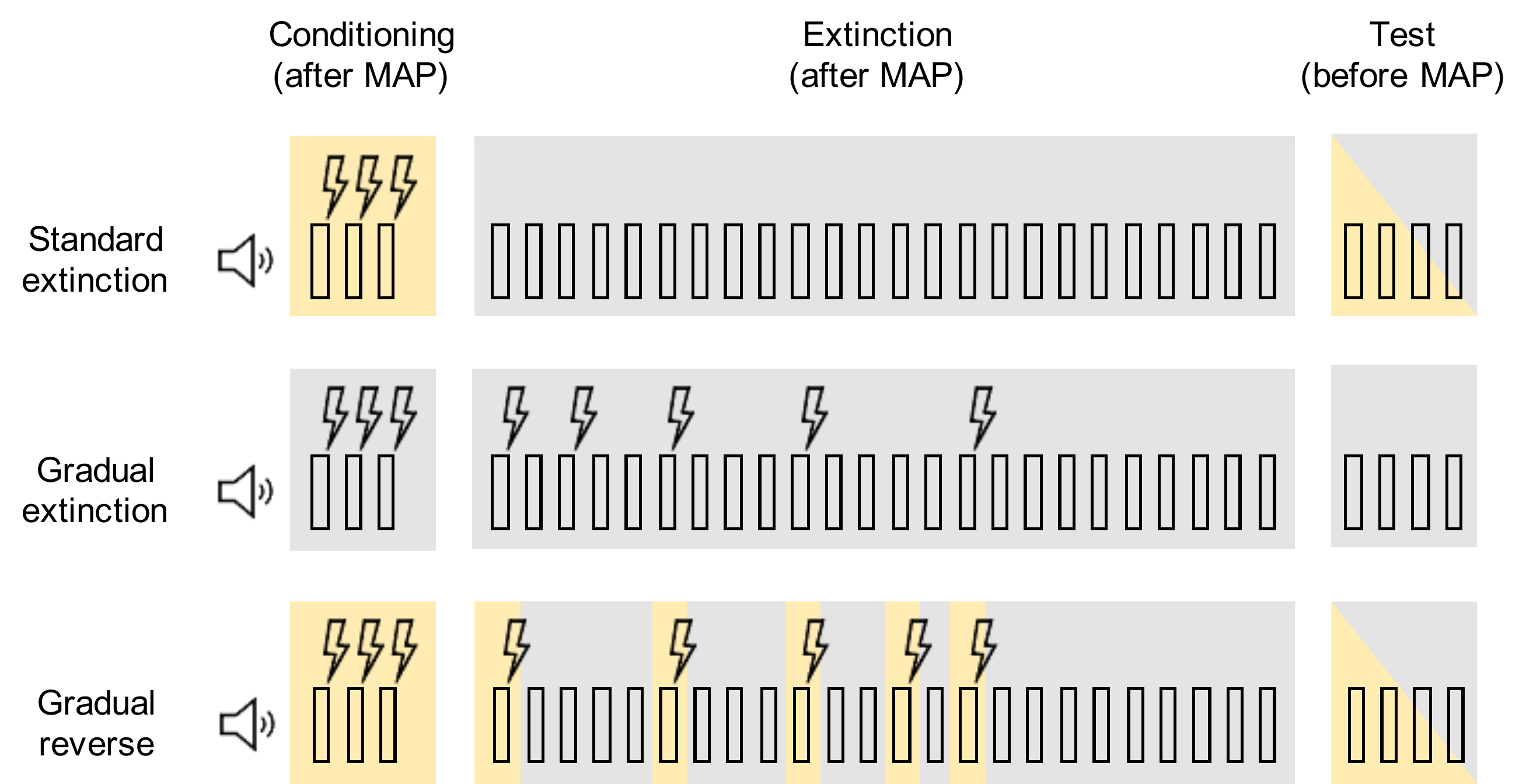}
\caption{\textbf{Trial sequences and model predictions for latent-cause assignments under the three extinction conditions.} Trial sequence in conditioning, extinction and test sessions: each rectangle represents the tone presentation in one trial; trials in which the tone co-terminated with a shock are marked by the 
lightning signs. Background shading colors indicate latent-cause assignments predicted by our model: a two-cause sequence for standard extinction and gradual reverse, and a one-cause sequence for gradual extinction. Splitting color at test indicates that both causes are likely. The possibility of a third, new, latent cause at test is not illustrated, as it is largely consistent across conditions and contributes little to freezing behavior. For conditioning and extinction sessions, the latent-cause assignments shown are those after each session, after collapsing to the mode of the posterior (MAP estimation; see text). For the test session, we show the probabilistic assignments before the MAP estimation as those generated the behavior measured. That is, all predicted assignments shown (both for conditioning and extinction sessions, and for the test session) are presumably what animals had in mind at test time, and therefore what governed freezing behavior during test.}
\label{fig:latent_cause_assignment}
\end{figure}

In addition to explaining the return of fear, the latent-cause inference framework prescribes a solution for effective extinction of the original fear association: instead of abruptly ``cutting off'' the pairing between tone and shock, which encourages the animals to infer a new latent cause, gradually decreasing their co-occurrence will make animals more likely to infer that the old cause is still active during extinction, but with decreasing tendency to generate shocks. To demonstrate this, Gershman and colleagues \cite{gershman2013gradual} conducted two \emph{gradual extinction} experiments (see Figure \ref{fig:exp_design} for experimental design), and indeed they observed reduced fear responses in both a spontaneous recovery test and a reinstatement test. In both experiments, they contrasted this condition with both a standard extinction condition and a \emph{gradual reverse} condition. In the latter, instead of gradually decreasing the frequency of the shock, they gradually increased it, while keeping the total number of shocks the same. Despite the surface similarity between the two gradual conditions (only two trials were different, though this difference markedly changed the trend of shock frequency from decreasing to increasing; Figure \ref{fig:latent_cause_assignment}), extinction was far less successful in preventing the return of fear in the gradual reverse condition, supporting the idea that abrupt changes encourage the inference of new latent causes. Since then, the effectiveness of gradual extinction has also been replicated in human participants \cite{shiban2015gradual}. There is also evidence of effective extinction in other experimental paradigms that create a gradual transition between the consistent presentation of shocks (during conditioning) and the absence of shocks. For example, occasional reinforcements during extinction eliminated spontaneous recovery in humans \cite{thompson2018enhancing, culver2018building}, and replacing the strong footshock used in conditioning with weak shocks during extinction (i.e., deconditioning) helped reduce both renewal and spontaneous recovery effects in rodents \cite{popik2020shifting}.

Conceptually, these effects can be explained by the latent-cause inference framework \cite{gershman2010context}; in fact, the experiments done by Gershman and colleagues \cite{gershman2013gradual} were inspired by this theoretical framework and aimed to test its predictions. However, it turned out that the original latent-cause inference model (presented in detail, e.g., in \cite{gershman2012exploring}), was not able to explain the empirical observations, in particular the difference between gradual extinction and gradual reverse conditions (S. Gershman, personal communication). The lack of a satisfactory computational account limited the conclusions that could be drawn from the original work, and its potential to inspire future investigation. 

To address this gap, here we describe a latent-cause model with additional assumptions that captures the pattern of the original empirical results. We show the model's predictions through simulations, and demonstrate the necessity of each model assumption by comparing with alternative models. Indeed, it is not trivial for the model to generate predictions that match animals' behavior across all three extinction conditions, and several important modeling assumptions are needed. These include assumptions on animals' prior belief over latent causes, the way in which they learn about the statistics of each latent cause, how animals reduce uncertainty in their inference process, and animals' response tendencies. We note that even with these assumptions, our model can only predict the comparative behavioral patterns between conditions, but not the exact freezing rates in each condition. We discuss potential reasons for this discrepancy. Although the original empirical results we model here may benefit from additional examination to show its generality, we propose that understanding the potential computational underpinnings of these findings can advance our understanding of extinction processes and how they can be made more effective. More broadly, this work provides insights into the potential mechanisms that support animal learning and inference, and generates new predictions to be tested in future experiments.

\section{Methods}

\subsection{The latent cause inference model}

We use a latent-cause inference model to explain the effectiveness (in terms of degree of return-of-fear) of standard extinction, gradual extinction and gradual reverse procedures, in both spontaneous recovery and reinstatement experiments. The model describes how animals infer the active latent cause on each trial by combining a prior belief with current observations, and how animals learn about the statistics of each latent cause. We assume such inference and learning were active throughout the entire experiment (including test sessions; see Results for experiment details). Specifically, as detailed below, the model we found to account for the experimental results uses the distance-dependent Chinese restaurant process as the prior, the Rescorla-Wagner update rule to learn about ``generative strength'' (i.e., the likelihood for each latent cause to generate observations, as detailed below), and assumes that animals approximate the (intractable) Bayesian inference process by collapsing their belief to the posterior mode between sessions. Additionally, to compare model predictions to empirical measurements, we make assumptions about how the prediction of shock maps to freezing behavior. We now describe each part of the model.

\subsubsection{Prior: distance-dependent Chinese restaurant process}

We assume that the animals' prior belief over latent causes takes the form of a distance-dependent Chinese restaurant process \cite{blei2011distance}, a variant of the Chinese restaurant process (CRP) infinite mixture-model prior \cite{aldous1985exchangeability}.

The standard CRP describes a categorization process with an {\it a priori} unlimited number of categories, whereby a new trial is more likely to be generated by a latent cause (category) that has generated more trials in the past. Specifically, the prior probability of an old cause generating the current trial is proportional to the number of trials this cause has already generated; the probability of the next trial being generated by a completely new latent cause is proportional to a fixed concentration parameter $\alpha$. Denoting the active latent cause on trial $i$ by $c_i$, the prior probability distribution over $c_i$ is thus:
\begin{equation*}
P(c_i = c|c_{1:i-1}) =
\begin{cases}
\frac{1}{i-1+\alpha} \sum_{j<i} \delta(c_j,c) &\text{ ($c$ is an old cause)}\\
\frac{1}{i-1+\alpha}\ \alpha &\text{ ($c$ is a new cause)}
\end{cases}
\end{equation*}
where $\delta(x,y)$ is the Kronecker delta function: $\delta(x,y)=1$ if $x=y$; otherwise, $\delta(x,y)=0$. Thus, $\delta(c_j,c)$ denotes whether the current cause $c$ is the same one that generated trial $j$. $\frac{1}{i-1+\alpha}$ is the normalization constant for this distribution. This distribution is exchangeable, meaning that it results in the same prior distribution over latent causes regardless of the order of trials.

Because trial order is important in the task we model, we used the the distance-dependent Chinese restaurant process (ddCRP), in which more distant experience contributes less to current inference through a decay function applied to the trial count. Specifically, we compute distance over time, using an exponential function with slope $k$. Since exponential decay can be arbitrarily close to zero (that is, an old latent cause not experienced for a long time can have a close-to-zero prior probability), departing from the classic ddCRP, we also add a baseline probability $b$ to all old latent causes. This corresponds to an assumption that any old latent cause has a non-negligible prior probability of becoming active again. The equations then become:
\begin{equation}
P(c_i = c|c_{1:i-1}) \propto 
\begin{cases}
\sum_{j<i} e^{-k (t_i-t_j)}\delta(c_j,c)  + b &\text{ ($c$ is an old cause)}\\
\alpha &\text{ ($c$ is a new cause)}
\end{cases}
\label{eq:prior}
\end{equation}
where $t_i$ is the time of trial $i$. The normalization constant for this distribution can be calculated by summing over the probability of all old causes and the new cause.

\subsubsection{Likelihood: Rescorla-Wagner learning}

We denote the generative strength of each observation $x$ for latent cause $c$ by $V(x|c)$, which describes the tendency of latent cause $c$ to generate $x$. Using the simplest and most widely-used learning rule in animal conditioning, i.e., the Rescorla-Wagner learning rule \cite{rescorla1972theory}, we assume that $V$ is updated for the currently active latent cause $c_i$ based on the observation of $x$:
$$
\Delta V(x|c_i) = \eta (x_i - V(x|c_i))
$$
where $\eta$ is the learning rate, and $x_i$ is the observation of $x$ on trial $i$ ($x_i=1 \text{ or } 0$ means $x$ is present or absent, respectively, on that trial).

To model fear extinction, we consider two types of observations: tone and shock. Given the different {\it a priori} prevalence of such stimuli in the animals' natural environment, we assume different initial values for the generative strengths for new latent causes: $V_0(\text{tone}) = 0.5$ and $V_0(\text{shock}) = 0.05$. We also assume a higher learning rate $\eta_\text{shock}$ for shocks (when shocks are present) considering their high motivational valence. These assumptions are important for the pattern of results, but the specific numeric values were not chosen through formal optimization or model-fitting.

We use the generative strength $V$ as the proxy for the animal's estimated probability of observing the tone or shock, i.e., the likelihood of the corresponding observation on that trial given a latent cause:
\begin{equation}
    P(x|c_i) = V(x|c_i).
    \label{eq:lik}
\end{equation}

\subsubsection{Update: Exact inference, with collapse of belief distribution between sessions}

We assume that animals perform Bayesian inference during each session. That is, they combine the prior probability and likelihood of both observations (assumed to be independently generated given the latent cause) to calculate the posterior belief distribution over the active latent cause:
\begin{equation*}
    P(c_i|\mathbf{x},c_{1:i-1}) \propto P(c_i|c_{1:i-1}) \prod_{x \in \mathbf{x}}P(x|c_i).
\end{equation*}
Here, the first term on the right-hand side is the ddCRP prior from above (Equation \ref{eq:prior}) and the second term is the likelihood of the current observations (Equation \ref{eq:lik} above). $\mathbf{x}$ denotes the combination of both observations (tone and shock). In this way, the animal maintains a probability distribution over each of the latent causes being active on each trial, updating this distribution as trials unfold.

Between experimental sessions, however, we assume that animals collapse their posterior belief distribution to its mode, i.e., a maximum {\it a posteriori} (MAP) estimation. In other words, we assume that animals do not maintain uncertainty over what latent cause was responsible for what observation in the previous session; instead, they ``pick'' the most likely sequence of latent causes for the past session, moving forward to the next session with only this deterministic assignment of trials to latent causes as the prior. Note that this is not a technical choice for faster model simulation, but an important modeling assumption for predicting the difference between gradual extinction and gradual reverse conditions (see \ref{sec:MAP}).

\subsubsection{Mapping prediction of shock to freezing behavior}

On each trial, as animals hear the tone and before the observation of a shock (on reinforced trials), we assume that animals use their current estimate of latent-cause assignment to predict how likely a shock is to occur on that trial, and thus decide whether or not to freeze in anticipation. The estimated shock probability is calculated by marginalizing over all possible latent causes:
$$
P(\text{shock}|\text{tone}) = \sum_{c_i}P(\text{shock}|c_i)P(c_i|\text{tone},c_{1:i-1}).
$$
Here, the last term is calculated using Bayes rule:
$$
P(c_i|\text{tone},c_{1:i-1}) \propto P(c_i|c_{1:i-1})P(\text{tone}|c_i).
$$

In mapping the animal's prediction of shock probability to freezing behavior, we make two more assumptions. First, we assume a non-zero baseline freezing rate (denoted by $p_0$): animals do not freeze in their natural environments; if, however, the animal becomes aware of shocks through experimental manipulations and anticipates them, empirical findings suggest that animals will show some baseline level of freezing behavior \cite{jacobs2010accurate}, regardless of how unlikely the shock is. In addition, for simplicity, we assume that freezing probability is proportional to the predicted shock probability:
$$
P(\text{freezing}) = (1-p_0) * P(\text{shock}|\text{tone}) + p_0.
$$

We also consider the locally perseverative nature of animals' behavior \cite{miller2019habits,lau2005dynamic}, and assume that there is a chance of $p_r$ that the animal will exhibit the same behavior as in last trial, regardless of its current prediction for the presence or absence of shock.

\subsection{Model simulations}

We used the following parameter values for simulating the model: $\alpha=0.2, k = 0.1, b=0.1, \eta=0.2, \eta_\text{shock}=0.4, V_0(\text{tone}) = 0.5, V_0(\text{shock}) = 0.05, p_0=0.2, p_r=0.7$. These values were initially set based on parameter values used in similar previous studies ($\alpha, \eta$) or a best first guess given our knowledge of animal behavior and the role of each parameter in the model ($k, b, \eta_\text{shock}, V_0(\text{tone}), V_0(\text{shock}), p_0, p_r$). We then fine-tuned the parameters with a grid search around that starting point, to best match empirical results. We did not perform likelihood-optimizing model-fitting because it is computationally prohibitive for this model. However, we note that simulation results were consistent across a range of parameter values. We also explored how model predictions depend on the change in parameter values (Supplementary Figure \ref{fig:newp}). Because it is computationally intractable to compute the full posterior distribution analytically, we used the particle filter algorithm as in \cite{gershman2010context} to approximate the posterior distribution with 10,000 particles.

We simulated the model and compared its predictions with the empirical results for the main experiments \cite{gershman2013gradual} (in Results) and a few additional experiments (in Discussion). The experimental protocols used in these experiments and our simulations, are described in detail in the following sections.

\section{Results}

In the following, we describe behavioral predictions of the model with all the assumptions described above (distance-dependent prior on cause assignment, Rescorla-Wagner rule for learning the generative strength of stimuli, MAP estimation between sessions, direct mapping from shock prediction to freezing behavior), and compare them to behavioral results in Gershman et al. \cite{gershman2013gradual}. We then turn to evaluating the necessity of each assumption by comparing with alternative models.

\subsection{Experimental measures and modeling goals}

We simulated the model and obtained its prediction for the three extinction conditions: standard extinction, gradual extinction, and gradual reverse. We considered two types of test (as in \cite{gershman2013gradual}): spontaneous recovery and reinstatement. Here, we first describe the experimental conditions in brief (see \cite{gershman2013gradual} for details on experimental design and subjects), as well as the goals of our modeling.

In the experiment, each rat completed one of the experimental conditions (Figure \ref{fig:exp_design}). All experiments started with a conditioning session (3 trials of a 20s tone, co-terminated with a 0.5s foot-shock). 24 hours later, animals underwent an extinction session (24 trials of the same tone) in one of three ways (Figure \ref{fig:latent_cause_assignment}): in standard extinction, the tone was presented alone in all trials; in gradual extinction, the tone co-terminated with a shock on trials 1, 3, 6, 10 and 15, and no shocks on other trials; in gradual reverse, the tone co-terminated with a shock on trials 1, 6, 10, 13 and 15, and no shocks on other trials. In this way, in gradual extinction, shocks gradually became less frequent, whereas in gradual reverse, shocks were made to be more frequent as the extinction session proceeded. In all cases, the last 9 trials of the extinction session did not involve shocks, to allow comparable extinction of freezing behavior before the subsequent test.

Animals were then tested for their fear response to the tone, measured by the percentage of time they spent freezing during the tone. In the spontaneous recovery experiment, 24 hours after the extinction session, the animals first underwent a so-called long-term memory test to test the extinction memory (with 4 trials of the tone alone); 30 days later, they were tested again for spontaneous recovery (4 trials of the tone alone). In the reinstatement experiment, 24 hours after the extinction session, animals experienced 2 unsignaled shocks without the tone. After another 24 hours, they were tested with 4 trials of the tone alone.

Our simulations focused on model predictions for (1) latent-cause assignment throughout the experiment; (2) animals' freezing rate in the last 4 test trials. Our goals were to demonstrate the latent-cause assignments that support animals' behavior under each extinction procedure, and more importantly, explain the qualitative differences between the three procedures in terms of their effectiveness in preventing the return-of-fear: gradual extinction being the most effective (i.e., lowest freezing rate at test, compared to the end of extinction), in comparison to standard extinction and gradual reverse. We note here, and return to this point in the Discussion, that while we strove to predict behavior throughout the experiment, due to large variability between individual rats, the behavioral pattern during extinction was hard to discern. We therefore did not attempt to quantitatively predict trial-by-trial behavior, or to fit the behavior of individual animals by using different parameter values for each animal. Instead, we focused on the average qualitative pattern of results at test, which meaningfully separated the different extinction procedures in terms of effectiveness. 

\subsection{Latent-cause assignment}

\vspace*{-4pt}

\begin{figure}
\centering
\includegraphics[width=\textwidth]{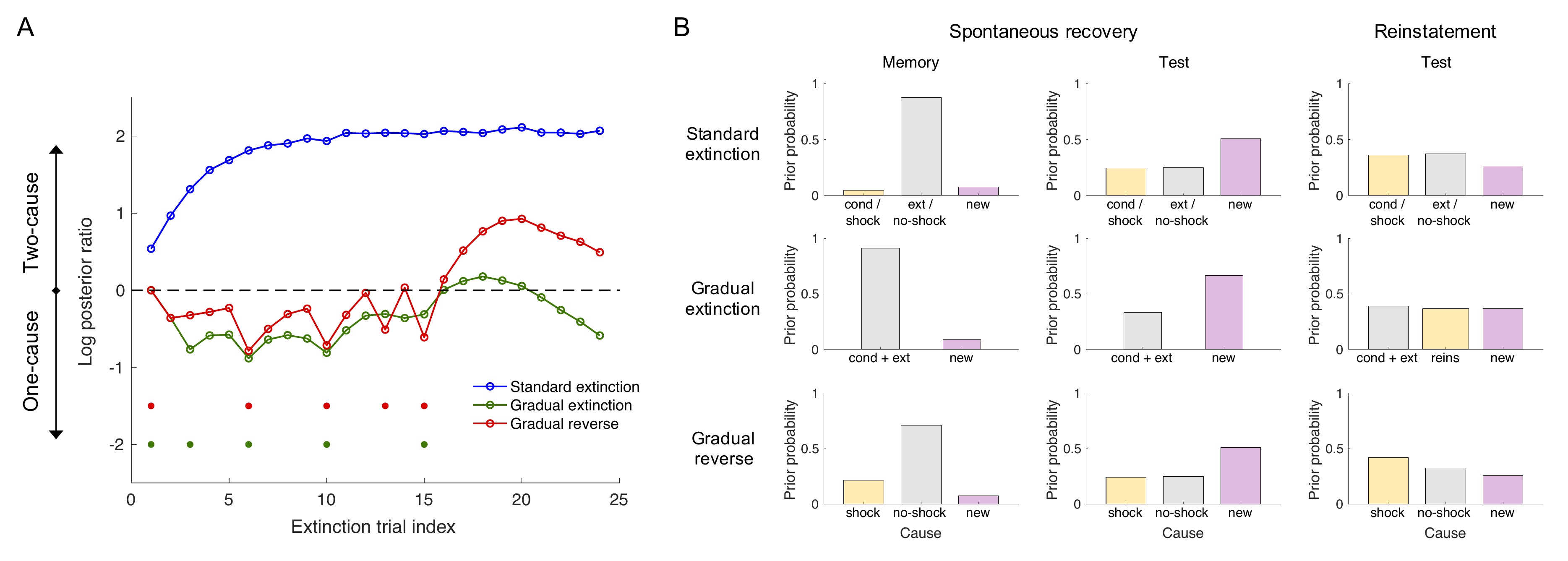}
\caption{\textbf{Model prediction for latent-cause probabilities.} \textbf{(A) Comparison between a two-cause sequence and one-cause sequence during extinction}. Log posterior ratio greater than (less than) zero indicates the dominance of the two-cause (one-cause) sequence in inference; the dashed line at zero represents the two types of cause sequence being equally likely. At the end of the extinction session, the two-cause sequence is more likely than the one-cause sequence in standard extinction and gradual reverse; in contrast, the one-cause sequence is more likely in gradual extinction. Here, ``two-cause'' sequence corresponds to the assignment of all shock trials to one cause, and all non-shock trials to another; ``one-cause'' sequence corresponds to all conditioning and extinction trials being generated by the same cause. \textbf{(B) Prior probability of latent causes in the first trial of the long-term memory and test sessions}. Left and middle columns: long-term memory and test sessions in the spontaneous recovery experiment; right column: test session in the reinstatement experiment. Top, middle and bottom rows: standard extinction, gradual extinction, and gradual reverse conditions, respectively. Causes are labeled based on what types of past trials they have generated: conditioning (cond), extinction (ext), reinstatement (reins), shock or no-shock. They are color-coded as in Figure \ref{fig:latent_cause_assignment}: yellow indicates a ``dangerous cause'', grey indicates a ``safe cause'', and purple indicates a new cause (with minimal prediction of shock, $V_0(\text{shock})=0.05$). } 
\label{fig:pcause}
\end{figure}

Figure \ref{fig:latent_cause_assignment} illustrates the model's assignment of trials to latent causes for the three extinction conditions. In this schematic, assignments in conditioning and extinction sessions are the deterministic results after the post-session MAP estimation (see Figure \ref{fig:pcause}A for the probabilistic assignments during the extinction session); these are the assignments that influence behavior at the test session, which is the focus of our interest. We also combine here the two types of test (spontaneous recovery and reinstatement) because their latent-cause assignments are similar (see Figure \ref{fig:pcause}B for latent-cause probability in each test). In standard extinction, two different latent causes are inferred for the conditioning session (all shock trials) and the extinction session (all no-shock trials); at test, both causes are likely (due to either reminder shocks or passage of time, both elevating the probability of the initial conditioning-session latent cause). In gradual extinction, because of the gradual reduction of shock probability, conditioning and extinction sessions are assigned to the same latent cause, as are the test trials. In gradual reverse, in contrast, all shock trials throughout conditioning and extinction are assigned to one latent cause, whereas all no-shock trials are assigned to a different latent cause; then, at test, both causes are likely.

\subsection{Prediction of freezing behavior}

\begin{figure}
\centering
\includegraphics[width=0.6\textwidth]{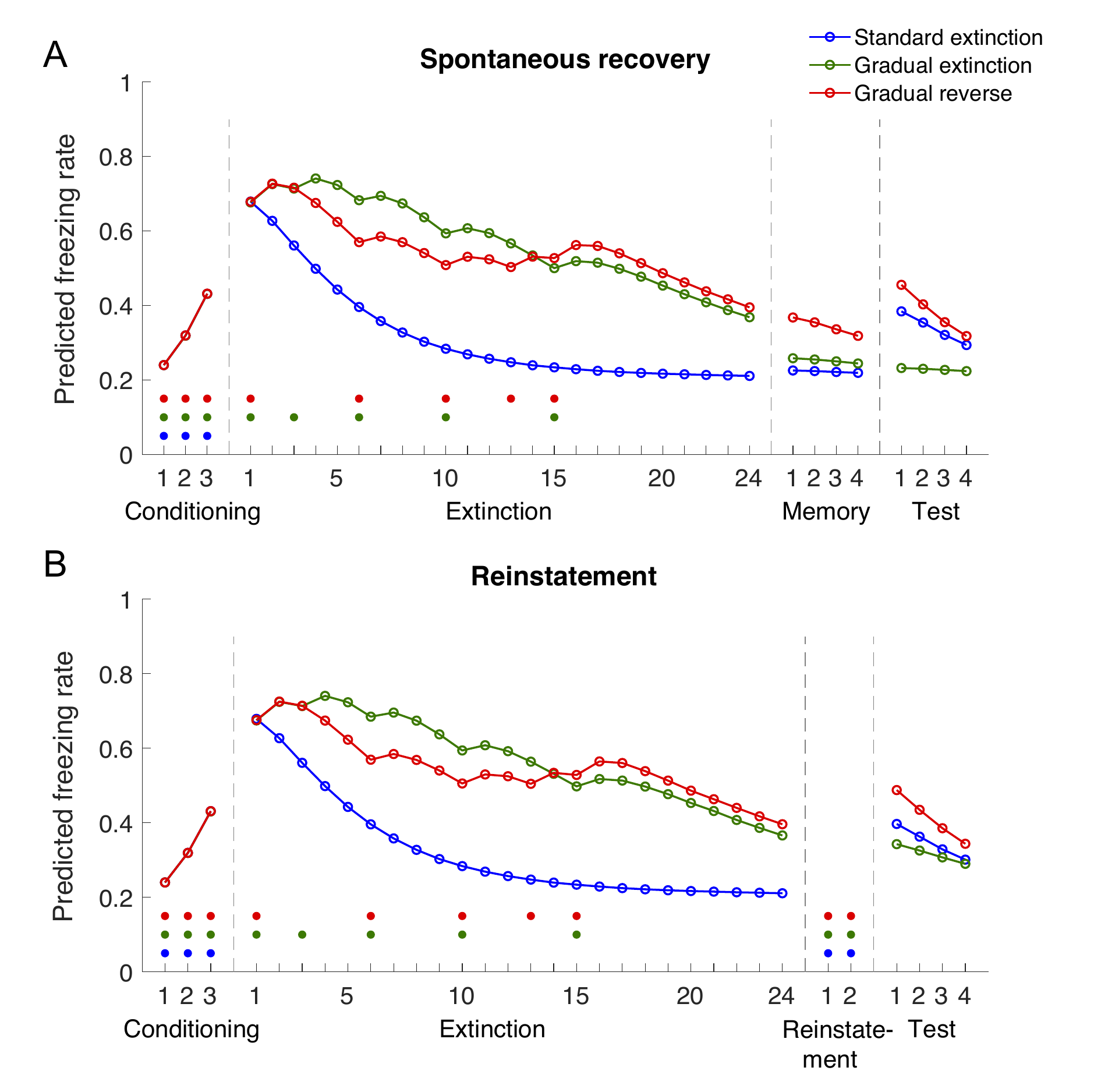}
\caption{\textbf{Model prediction of freezing rate for (A) spontaneous recovery and (B) reinstatement experiments.} Under all three extinction conditions, freezing rate increases during conditioning, and decreases during extinction. In the beginning of test session, however, freezing rate jumps back up for standard extinction and gradual reverse conditions, but remains low in gradual extinction. Note that the model predicts freezing rate upon tone presentation, before the actual delivery (or absence) of shock. Dots at the bottom of the plots indicate shocks in the corresponding trials, color-coded based on the extinction conditions. Dashed gray vertical lines indicate session boundaries (in practice: at least 24h gap).}
\label{fig:pshock_curve}
\end{figure}

Figure \ref{fig:pshock_curve} shows model prediction for freezing rate across experiment sessions. For all three extinction conditions, freezing rate increases during conditioning, decreases during extinction (more so in standard extinction, with no shocks in the extinction session), and continues to decrease in the long-term memory test (24 hours after extinction in the spontaneous recovery experiment). However, in the test session (both spontaneous recovery and reinstatement experiments), the predictions for the three conditions diverge, due to the distinct latent-cause assignments: in standard extinction and gradual reverse conditions, freezing rate increases compared to the end of extinction, showing the return of fear; in the gradual extinction condition, it continues to decrease both in the spontaneous recovery and the reinstatement tests.

\begin{figure}[h!]
\centering
\includegraphics[width=0.8\textwidth]{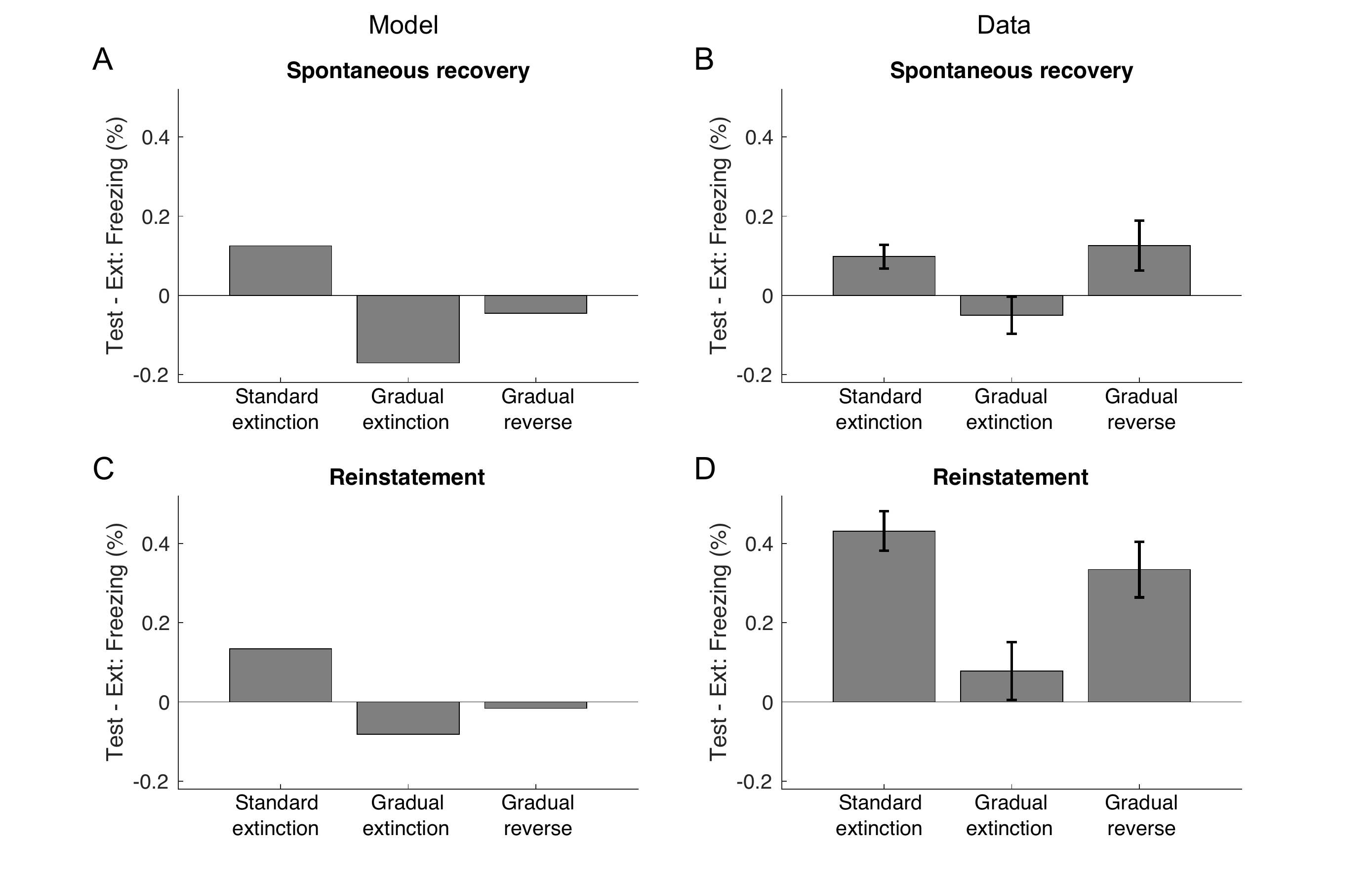}
\caption{\textbf{Impact of different forms of extinction on return of fear: model predictions (A,C) are qualitatively consistent with empirical results (B,D) in both experiments.} Model simulations correctly predict that standard extinction leads to the greatest return of fear, whereas gradual extinction is the most effective in permanently reducing fear, across both spontaneous recovery and reinstatement tests. The effectiveness of extinction is calculated as the difference in freezing rate between the four test trials and the last four extinction trials (all no-shock trials). Panels B and D are reproduced from \cite{gershman2013gradual}.}
\label{fig:test_effect}
\end{figure}

Figure \ref{fig:test_effect}A and \ref{fig:test_effect}B summarize model predictions on the difference in freezing rate between the four test trials and the last four extinction trials, for comparison with the qualitative pattern in the empirical results (Figure \ref{fig:test_effect}C and \ref{fig:test_effect}D). According to the model, for both spontaneous recovery and reinstatement tests, fear response reduces the most in gradual extinction, followed by gradual reverse; fear response at test increases in standard extinction. We note here (and discuss in more detail below) the clear discrepancies between our simulation results and the empirical results, where in simulation gradual reverse does not result in overall increase in fear at test as compared to the end of extinction. Nevertheless, the model predictions are qualitatively consistent with the empirical findings, illustrating the success of the current model in explaining the relative pattern seen in the experimental results.

\subsection{Necessity of model assumptions}
\label{necessity}

The model described above was tailored to explain the behavioral patterns in the data by adding assumptions as needed where the original CRP model \cite{gershman2010context} did not suffice. We now turn to discussing these model assumptions and demonstrating their necessity. We do so by comparing the predictions of the current model (referred to as ``main'' model below) to reduced models that do not include these assumptions. Since we did not fit the parameters of the main model to the data using statistical techniques, we do not present formal model comparisons, but rather focus on the qualitative patterns in the data.

\begin{figure}
\centering
\includegraphics[width=\textwidth]{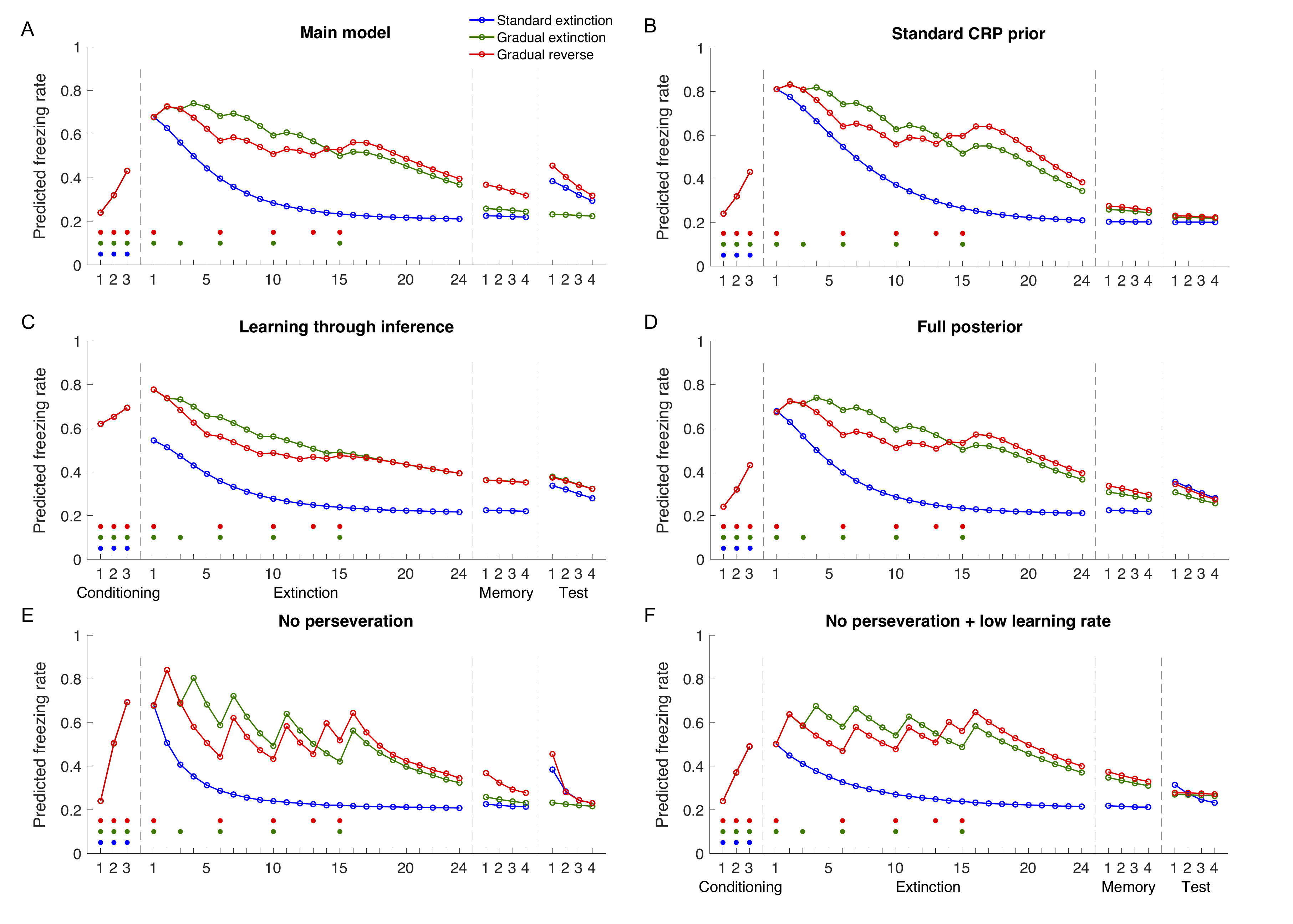}
\caption{\textbf{Simulation of alternative models demonstrates the necessity of assumptions in the main model.} Each alternative model differs from the main model in only one assumption; model simulations were conducted with the same parameter values, except for the alternative assumptions, as noted below. Shown are simulations of the spontaneous recovery experiment; results are consistent for the reinstatement experiment. \textbf{(A)} The main model, same as Figure \ref{fig:pshock_curve}A. \textbf{(B)} Standard CRP prior (without distance dependence; $k=0$ in the ddCRP) assigns trials in the memory test (24 hours later) and spontaneous recovery test (30 days later) to the same latent causes, and thus shows no spontaneous return-of-fear after 30 days for any of the conditions. \textbf{(C)} Learning generative strengths through inference (i.e., Bayesian inference on Bernoulli probabilities for tone and shock, instead of Rescorla-Wagner learning) predicts the same latent-cause assignments and expected shock probability for gradual extinction and gradual reverse conditions, and thus cannot explain their difference in fear responses during test. In this alternative model, we replaced the Rescorla-Wagner learning parameters $\eta, \eta_\text{shock}, V_0(\text{tone}), V_0(\text{shock})$ with parameters for pseudo-counts: $N(\text{tone})=0.5, N(\text{no-tone})=0.5, N(\text{shock})=0.95, N(\text{no-shock})=0.05$. These pseudo-counts determine the prior probabilities of tone and shock (each denoted by $x$): $p(x)=N(\text{x}) / \left(N(\text{x})+N(\text{no-x})\right)$, reflecting a similar asymmetry in the \textit{a priori} prevalence of tone and shock as in the main model. \textbf{(D)} Keeping the full posterior distribution over latent causes across sessions (no MAP estimation) predicts minimal return-of-fear in the gradual reverse condition. \textbf{(E)} No perseveration ($p_r=0$) leads to over-sensitivity to shock/no-shock experience during extinction which was not seen in the empirical data, as well as rapid reduction in return-of-fear during test. \textbf{(F)} Replacing the perseveration assumption with lower learning rates to stabilize responding across trials ($p_r=0, \eta=0.1, \eta_\text{shock}=0.2$) fails to predict the return-of-fear effect in the gradual reverse condition.}
\label{fig:alternatives}
\end{figure}

\subsubsection{Distance-dependent CRP: spontaneous recovery depends on test delay}

Experiments suggest that spontaneous recovery of fear response increases with the delay between extinction and test \cite{pavlov1927conditioned, rescorla2004spontaneous}. For example, the standard extinction group in Gershman et al. \cite{gershman2013gradual} showed no significant return of fear in the memory test 24 hours after extinction (paired sample t-test: $t(15) = -1.80, p = .09$, in comparison to the last four extinction trials). In contrast, the rats showed significantly more freezing at test 30 days later, compared to the last four trials of extinction (paired sample t-test: $t(15) = 3.26, p < .01$). 

Our model captures the dependency of spontaneous recovery on delay duration using the distance-dependent CPR prior, similar to what was used in \cite{gershman2017computational}. According to this prior, the probability of an old latent cause being active depends on the time between its previous instances and the current trial. The closer they are in time, the more likely the old cause will be active again. As a result, the model predicts that the latent cause inferred for the extinction session is very likely to continue to be active in the memory test (Figure \ref{fig:pcause}B left column), as the memory test is much closer to the extinction session (24 hours apart) than to the conditioning session (48 hours). After 30 days, however, both conditioning and extinction sessions are similarly distant, so the model predicts that both the conditioning and extinction latent causes are equally likely to be active (Figure \ref{fig:pcause}B middle column; of course, a completely new latent cause is more likely in this case), resulting in an increased fear response in standard extinction and gradual reverse. In gradual extinction, since there has been only one latent cause throughout, this cause (with reduced generative strength of shock) is inferred to be active again during test, resulting in no return-of-fear above and beyond what was observed at the end of extinction.

An alternative model that uses the standard CRP prior with no distance dependence (and keeps all other assumptions and parameter values the same as the main model) produced latent-cause assignments that are independent of test time. It thus made similar predictions for the memory test and the spontaneous recovery test, failing to predict the spontaneous recovery of fear in standard extinction and gradual reverse conditions (Figure \ref{fig:alternatives}B). 

Note that we are not committed here to the specific form of the distance dependence and have not exhaustively tested other forms of recency-weighted or perseverative CRP priors (e.g., \cite{lloyd2013context}), or different decay functions. Our claim is only that an order-agnostic exchangeable prior distribution, as is commonly used in machine learning applications to categorization, would not accord with the empirical data.

\subsubsection{Rescorla-Wagner learning (recency-weighted estimates):  difference between gradual extinction and gradual reverse}

In the latent-cause inference framework, it is common to assume that the generative strength of observations is fixed (though unknown) for each latent cause \cite{gershman2010context, gershman2012exploring}. Under this assumption, the optimal estimator of the shock probability for each latent cause is the proportion of trials in which shocks appeared under that cause. This assumption is at odds with the behavioral differences between gradual extinction and gradual reverse, as both conditions had the same number of shocks and would therefore predict identical test behavior (Figure \ref{fig:alternatives}C).

To explain the behavioral difference between gradual extinction and gradual reverse, we thus found it necessary to assume a recency-weighted estimate of generative strength, such as given by the Rescorla-Wagner (RW) learning rule. Similar learning rules have been used in latent-cause inference models of associative learning and memory modification \cite{soto2014explaining, gershman2017computational}. The RW rule learns a dynamic shock probability through an error-correcting process that adjusts estimates proportionally to the error in predicting the current observation. A fixed learning rate results in over-weighing of recent experiences, effectively estimating the generative strength as an exponentially-decreasing average of previous experiences. This is normative if animals assume that the environment may change over time and track such changes. In our case, this allows them to treat differently a decrease in shock probability in gradual extinction versus an increase in gradual reverse. As a result, they make different latent cause assignments for gradual extinction and gradual reverse, and in turn, show distinct levels of fear responses at test.

\subsubsection{Collapsing uncertainty (MAP) between sessions: return-of-fear in gradual reverse}
\label{sec:MAP}

Latent-cause inference involves evaluating the likelihood of all possible cause sequences for the past trials -- from all trials being generated by one cause, to each trial being generated by its own unique cause, through any combination in between. Probabilistic inference means that the model does not commit to any of these assignment sequences; instead, all are likely during inference, and each is associated with some non-zero probability. In particular, in all three extinction conditions, both a one-cause sequence and a two-cause sequence are likely (Figure \ref{fig:pcause}A). While the two-cause sequence dominates in standard extinction, the assignments for gradual extinction and gradual reverse are more uncertain, starting with similar probabilities in both conditions and diverging through extinction. At the end of the extinction session, the one-cause sequence establishes some advantage over the two-cause sequence for gradual extinction, and the opposite is true for gradual reverse. The MAP assumption collapses this uncertainty over latent-cause sequences at the end of the extinction session, obtaining the deterministic (and different) assignments shown in Figure \ref{fig:latent_cause_assignment}. Specifically, the model commits to the one-cause assignment for gradual extinction, and the two-cause assignment for gradual reverse, accentuating the small differences between these conditions at the end of the extinction session. Similar collapsing of uncertainty has been applied in domains like perceptual decision-making (e.g. \cite{stocker2007bayesian}) to facilitate inference and decisions where estimating the full posterior distribution is computationally intractable.

We found the MAP assumption to be necessary for predicting the return-of-fear effect in gradual reverse. An alternative model that keeps the full distribution over latent causes throughout the experiment failed to predict the increase in freezing rate during test for gradual reverse (Figure \ref{fig:alternatives}D). This is because the two-cause sequence was deemed only somewhat more likely during test, which resulted in minimal return of fear. Similarly, because of the probabilistic latent cause assignments, the two-cause sequence had a non-zero probability in gradual extinction, resulting in a slight increase of freezing rate at test compared to the main model prediction.

\subsubsection{Perseveration: gradual change in behavior during extinction, and persistence of return-of-fear during test}

In the main model, we postulate that animals tend to repeat what they have been doing in past trials with probability $p_r$. Such perseverative behavior has been widely observed in perceptual and value-based choice tasks, for both animals and humans \cite{ito2009validation,gold2008relative,lee2005learning,akaishi2014autonomous}. In the current task, perseveration is important for predicting the persistence of return-of-fear during the test session. Without this assumption, fear responses will decrease rapidly in both standard extinction and gradual reverse conditions, as early as the second test trial (Figure \ref{fig:alternatives}E). This is due to animals' inference of latent causes: experiencing one trial without shock is sufficient to infer that the ``safe cause'' is active in the test session, and due to the distance-dependent CRP prior, this inference comes to dominate latent-cause assignments at test. However, such rapid reduction in freezing behavior was not seen behaviorally.

Simulations without perseveration also predicted freezing behavior that was overly sensitive to shock and no-shock experiences during extinction (Figure \ref{fig:alternatives}E): in gradual extinction and gradual reverse, simulated freezing rate jumped up following shocks and dropped following no-shocks. Examining animals' freezing rate during extinction (by re-scoring of the original videos using a convolutional neural network \cite{cai2020distinct}, as the middle trials during extinction session were not scored or reported in the original paper) suggested that behavior did not reflect trial-by-trial shock delivery or absence; on average, freezing rate noisily but gradually decreased through extinction session in all three conditions (not shown). An alternative explanation for the more gradual behavioral change in extinction can be a lower learning rate in updating the generative strengths of latent causes. However, reducing the learning rate interfered with latent-cause inference as it rendered animals less sensitive to shocks and thus biased them towards the one-cause assignment. An alternative model where we halved the learning rates predicted no return-of-fear in gradual reverse (Figure \ref{fig:alternatives}F). 

We therefore used the perseveration assumption to account for both rapid learning in fear conditioning and extinction, and gradual change in behavior. Additionally, we note that due to local perseveration, the main model predicted an increase in freezing rate at the beginning of the extinction session (compared to the end of conditioning session; Figure \ref{fig:alternatives}A), potentially related to the well-documented ``extinction burst'' phenomenon \cite{skinner1938behavior, cooper2020applied}.

\section{Discussion}

In this work, we use a latent-cause inference model to explain the differential effectiveness of gradual versus abrupt (standard) fear-extinction procedures. The model explains the return-of-fear effect commonly observed in standard extinction by presuming that animals infer a new state of the world during extinction and thus form a new association between tone and shock, as opposed to unlearning the original association. Similar ideas have been proposed both under similar statistical inference framework \cite{courville2006bayesian,gershman2010context} and through reinforcement-learning models with a state-classification mechanism \cite{redish2007reconciling}. This explanation also aligns with decades-old suggestions that extinction results in learning of a new safe association that competes with the original threat association, though does not override it \cite{bouton2004context}. The novelty of our work lies in formulating the inference and learning processes, demonstrating with quantitative simulation the differences between three extinction procedures, and verifying the necessity of various model assumptions. 

We show through model simulations that animals make distinct inferences of latent cause assignments under different extinction procedures: gradual extinction is the most effective at extinguishing the original fear association because both conditioning and extinction sessions are assigned to the same latent cause, and the gradual reduction in shock helps animals acquire a decreasing estimate of shock probability, leading to minimal return-of-fear during test. In contrast, the abrupt change in shock frequency in the other two procedures (complete absence of shocks in standard extinction; abrupt reduction in shock appearance, followed by increasing shock frequency in gradual reverse) results in the creation of a new ``safe'' latent cause to which all subsequent no-shock trials are assigned. This new latent cause protects the old ``dangerous'' latent cause from being updated by the no-shock experiences. As a result, the original fear association remains intact and can resurface at test. 

\subsection{Additional model assumptions}

We found that several additional model assumptions were needed to predict the behavioral findings: (1) a distance-dependent prior on latent-cause assignments that used the passage of time to determine distance; (2) learning the dynamics of the environment through a recency-weighted rule such as the Rescorla-Wagner learning rule; (3) reduction of uncertainty over the posterior distribution through MAP estimation between sessions; (4) behavioral perseveration. Without each of these assumptions, the model failed to replicate the higher freezing rates at test in standard extinction and gradual reverse conditions in comparison to gradual extinction. Most of these assumptions build on past models that have successfully explained a wide range of phenomena in animal (and human) learning and decision-making, as we discuss below. Moreover, the necessity of each assumption also suggests principles of animal learning mechanisms in Pavlovian tasks and beyond.

The distance-dependent CRP prior we used highlights the important role of time in latent-cause inference, especially when behavior is examined at different time intervals, or even on different time scales. We are not the first to propose such time-dependency. Gershman and colleagues \cite{gershman2017computational} used a similar distance-dependent CRP prior (with a power-law temporal kernel) to explain how spontaneous recovery depends on extinction-test interval, as well as why the effect of post-retrieval memory modification is sensitive to memory age. Additional empirical evidence for time-dependent inference process comes from work on extinction delay. Myers and colleagues \cite{myers2006different} found that varying the delay between acquisition and extinction affected the amount of return-of-fear: animals that experienced extinction trials 10 minutes or 1 hour after acquisition showed little or no subsequent spontaneous recovery, reinstatement or renewal effects; whereas those who had extinction trials 24 or 72 hours after acquisition showed strong return of fear effects. This effect of extinction delay cannot be explained by a latent-cause inference model without temporal dependency. The distance-dependent CRP prior can explain these findings as a result of the decreasing probability of the extinction trials being generated by the same latent cause as the acquisition trials when the delay between extinction and acquisition increases. Thus, with a shorter delay (10 min or 1h), the animal is more likely to classify the extinction trials into the same latent cause as the acquisition trials, which helps the successful unlearning of the shock probability, and thus prevents the return of fear.

The importance of time is also reflected in the recency-weighted learning rule we used. The Rescorla-Wagner learning rule has been widely used to model classical conditioning. It has also been used in latent-cause inference models to explain compound generalization in associative and causal learning \cite{soto2014explaining} and memory modification \cite{gershman2017computational}.

Together, the distance-dependent CRP prior and the Rescorla-Wagner learning rule imply specific beliefs that animals may have about the environment. Both modeling assumptions reflect the animal's inner model of the environment, i.e., whether they consider it as static or changing over time. According to the distance-dependent CRP prior, latent causes that were last active a long time ago are less likely to be active again, suggesting that the passage of time can lead to changes in the environment, making older causes less likely. Similarly, the Rescorla-Wagner learning rule over-weighs more recent experience, inherently capturing how the statistics of observations within a latent cause may change over time. Alternative assumptions (standard CRP prior and learning through exact Bayesian inference) are better suited to static environments. The advantage of the current model over alternatives provides evidence that animals are capable of acquiring rich dynamics in their learning environment, and that their inner model of the environment is in accord with the actual changing nature of naturalistic environments.

We can further consider a more general approach to modeling learning in a changing environment: deriving the normative learning rule based on the generative model. For example, the Kalman filter model \cite{kalman1960new} has been suggested as a model for estimating the mean and standard deviation of a Gaussian reward distribution under the assumption that it evolves over time\footnote{In fact, Rescorla-Wagner learning can be seen as a special case of the Kalman filter with a fixed Kalman gain and without tracking uncertainty \cite{gershman2015unifying}}. Another option, normative for environments in which the amount of reward changes at a constant rate, is to add a ``momentum'' term (calculated as the running average of recent prediction errors) to the Rescorla-Wagner learning rule \cite{eldar2016mood}.\footnote{We also tested an alternative model with momentum; the results were largely consistent with the RW learning rule. Thus, for simplicity, we used the more basic RW rule in the main model.} Future work can derive normative learning rules for the three extinction procedures in the current experiment, and test whether they account for behavior better than the Rescorla-Wagner learning rule we used.

\begin{figure}
\centering
\includegraphics[width=0.9\textwidth]{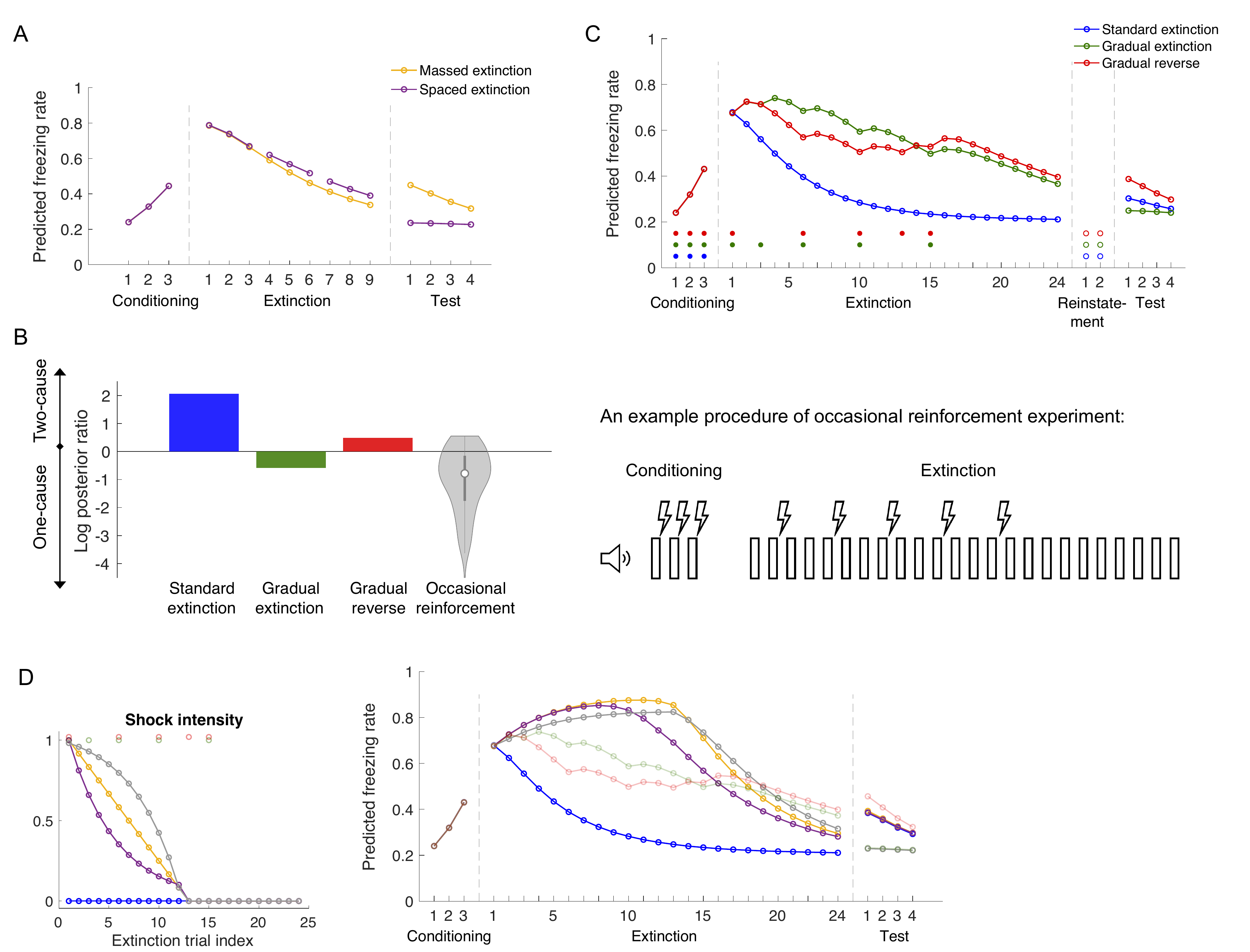}
\caption{\textbf{Simulations of additional experiments.} The same model and parameter values were used as in the main experiments, unless otherwise noted. \textbf{(A)} Spaced extinction is better than massed extinction at reducing long-term fear. We simulated the spaced and massed extinction procedures similar to \cite{urcelay2009spacing}, with spaced extinction comprised of 3 trials per session for 3 sessions in total and massed extinction comprised of 9 trials in a single session (all extinction trials involved presentation of the tone without a shock). Spontaneous recovery was tested after 30 days, and showed considerable return of fear for the massed condition, and markedly less fear in the spaced condition. In this simulation, $\alpha = 0.05$. Note that the lower $\alpha$ value used here indicated a lower tendency to infer new latent causes, which could be due to the animals' additional prior shaping experience in \cite{urcelay2009spacing}. \textbf{(B)} Occasional reinforcement during extinction favors the inference of one latent cause over two latent causes. To simulate the occasional reinforcement, five shocks were randomly placed in the first 15 trials in the extinction session, with the terminal 9 trials being no-shock trials; 100 different experiments (each with shocks randomly allocated to trials) were simulated. Left: log posterior ratio between a two-cause belief and a one-cause belief at the end of extinction session, for the three extinction procedures as in \cite{gershman2013gradual} and the occasional reinforcement procedure. Violin plot shows the median (dot), interquartile range (darker error bar), and 95\% confidence interval (lighter error bar) across the 100 simulations. Right: an example procedure of the occasional reinforcement experiment. \textbf{(C)} Low intensity reinstatement (modeled as no-shock trials) leads to minimal return of fear. The same experimental setup was used as in \cite{gershman2013gradual}, except that there was no shock in the two reinstatement trials (to approximate animals' perception of low intensity shocks). There was less return of fear at test compared to the original results (Figure \ref{fig:pshock_curve}B) in both standard extinction and gradual reverse procedures. \textbf{(D)} Gradually reducing shock intensity prevents return-of-fear. Left: shock intensity in several extinction procedures with reduced intensity over time (in gray, yellow, and purple curves; note that the three curves overlap at 0 in trials 13 to 24.). The three procedures reported in \cite{gershman2013gradual} are also shown as references (standard extinction: blue; gradual extinction: green; gradual reverse: red; all shocks in gradual extinction and gradual reverse were modeled with an intensity of 1, plotted with jitters for better visualization; no-shock trials were modeled with an intensity of 0, not plotted). Right: predicted freezing rate during conditioning, extinction and spontaneous recovery test (30 days later); color codes as in the left panel (note that blue, purple and yellow curves overlap at test). The procedure with the most gradual change in shock intensity at the beginning of extinction (gray curve, in contrast to yellow and purple curves) best prevents long-term return of fear. In this simulation, shock intensity was modeled as a continuous variable with Gaussian noise $N(0,\sigma^2)$ ($\sigma$ is an additional model parameter which was set to 0.43; all other parameter values were kept as in the original model).}
\label{fig:other_exp}
\end{figure}

Compared to the above two assumptions, the MAP assumption (collapse of the posterior to its mode between sessions) is found less often in the animal-learning literature. This assumption, nevertheless, also reflects the effect of the passage of time on inference, and can be construed as a result of memory consolidation (e.g., via replay of past events during sleep \cite{stickgold2005sleep}). Through consolidation, animals may revisit their learning experience in the previous session, and continue to update their belief over latent causes accordingly. Such update can reinforce the most probable latent-cause assignment, and eventually make it the only possibility, equivalent to a MAP estimate. Similar consolidation mechanisms have been proposed during long inter-trial intervals. For example, to explain memory modification, Gershman and colleagues \cite{gershman2017computational} introduced a ``rumination'' process taking place between the re-exposure of previous memory and the attempt to modify/extinguish it. Such rumination reinforces the dominant belief of returning to the past context, and facilitates memory modification. There is also extensive evidence on the superiority of spaced learning (with longer intervals between training examples, including overnight session boundaries) over massed learning, facilitated by memory consolidation \cite{smolen2016right}, in both animals and humans \cite{ebbinghaus1964memory, lattal1999trial, commins2003massed, cepeda2006distributed}. Specifically, Urcelay and colleagues \cite{urcelay2009spacing} found that spaced extinction trials (3 trials per session, for 3 sessions in total) led to less spontaneous recovery and renewal of the original memory than massed trials (9 trials in a single session). We replicated this result by simulating our model (albeit with a lower value of the concentration parameter $\alpha$ than used in other simulations; Figure \ref{fig:other_exp}A). In the model, the difference between the conditions is caused by the collapse of the posterior distribution between sessions: over a single long session, the inferred probability for the two-cause sequence increases throughout the session, eventually exceeding that of the one-cause sequence and leading to a two-cause belief. In contrast,  at the end of each short session of spaced extinction, the one-cause sequence is still more probable and is therefore ``committed'' through the MAP mechanism. This results in a one-cause belief at the end of extinction, and therefore animals show less return of fear under spaced extinction. Similar results have also been found in deconditioning experiments \cite{popik2020shifting} where a single long deconditioning session leads to return-of-fear but multiple short ones do not.

From a normative perspective, MAP estimation may be justified to allow inference in complex scenarios where the full distribution over latent causes is computationally intractable. Because of the combinatorial explosion of possible latent-cause sequences as more trials are experienced, any experience that unfolds over a sequence of events quickly becomes intractable. It is thus reasonable to assume that the brain collapses uncertainty over previous inference periodically (perhaps facilitated by large gaps in experience, replay of experiences, and memory consolidation), to be able to continue building on past knowledge without necessarily maintaining all of it. Indeed, Stocker and colleagues \cite{stocker2007bayesian,luu2018post} have shown that biases in human decision making can be explained by postulating that people collapse parts of their posterior distribution between sequential decisions, essentially discarding beliefs that are inconsistent with actions they have already made. Our MAP assumption is a similar form of commitment to the most likely past beliefs over others. 

In a different but related neural network model of memory modification \cite{osan2011mismatch, popik2020shifting}, attractors are used to model memory traces, similar to the latent causes in the current framework. The probabilistic retrieval and modification of existing attractors, and the formation of new attractors are similar to the inference and learning on latent causes. In the network model, learning results in stored attractors in the network, which is conceptually similarly to the deterministic latent-cause assignments as a result of MAP approximation in the current model.

The MAP assumption provides novel theoretical predictions that can be tested in future experiments. A direct test of the memory consolidation account of the MAP estimation would be to reduce the time between extinction and test in the reinstatement experiment. For example, conducting extinction, reinstatement and test all within one day or even one session. If memory consolidation indeed facilitates the collapse of the posterior distribution and strengthening of one interpretation of past events, animals with no or less time between extinction and test should maintain the probabilistic belief at test, and thus show less freezing in the gradual reverse condition. Our model does not differentiate between an abrupt collapse of the posterior distribution and a gradual reduction of posterior uncertainty. To further examine the process of uncertainty reduction, future work can manipulate the time interval between extinction and test, and examine how freezing rate at test changes as a function of such time interval.

The reduction of uncertainty due to the MAP estimation may seem at odds with the increase in uncertainty over time that results from the dynamic generative model as discussed earlier. However, it is worth noting that the MAP reduction of uncertainty affects categorization of past experiences, whereas the increase in uncertainty pertain to predicting future experience. On the one hand, the collapse of the posterior may suggest the limitation of animals' representation of the world; perhaps animals are able to learn a rich representation, but fail to maintain uncertainty about this representation over long periods of time. On the other hand, MAP estimation can also be seen as a method by which animals (and humans) build concise models of the world.

Last but not least, the perseveration assumption suggests a value-free habitual system (i.e., animals tend to repeat their previous behavior regardless of their prediction in the current scenario, as posited by Miller and colleagues \cite{miller2019habits}) that exists alongside a model-based system corresponding to the latent cause inference mechanism in the current model. Such a dichotomy has been widely observed in animal learning and decision-making. The fact that animals' behavior does not fully reflect their learned world model or even stimulus values underscores the importance of accounting for common habitual behavior (e.g., perseveration, side-bias, etc.) in modeling so that the habitual part of behavior will not mask the rich learning mechanisms.

\subsection{Related empirical results}

In this work, we provided a quantitative and theoretical account explaining why gradual extinction is more effective in permanently reducing fear than standard extinction and gradual reverse procedures. We focused specifically on predicting the empirical findings in Gershman et al. \cite{gershman2013gradual}; however, it is worth noting that there is other evidence on the effectiveness of gradual extinction.

In a fear-extinction experiment with human participants \cite{shiban2015gradual}, gradual extinction was shown to prevent the return of fear better than standard extinction, as measured by startle response (although there were no effects for contingency rating or skin-conductance response). Additional evidence comes from occasional reinforcement experiments: having occasional reinforced trials during extinction (effectively reducing the shock probability more gradually compared to standard extinction procedure) has been shown to eliminate spontaneous recovery in humans \cite{thompson2018enhancing, culver2018building}. Similarly, in appetitive conditioning experiments, occasional presentations of reinforcement in extinction slowed the re-acquisition of conditioned responses \cite{bouton2004occasional, van2015effects}, suggesting unlearning during extinction rather than learning of a competing association that would allow rapid relearning by activating the original association. We simulated the effect of receiving occasional reinforcement during extinction with our model (Figure \ref{fig:other_exp}B): according to the model, by the end of extinction, animals would be more likely to believe that conditioning and extinction sessions were generated from the same cause, rather than two distinct causes; therefore, they would show less return of fear.

We note that seemingly opposing evidence was observed in a reinstatement experiment by Rescorla \cite{rescorla1975reinstatement}, where gradual extinction was less effective than standard extinction in preventing the reinstatement effect. However, the reminder shocks used in this experiment were very weak (compared to other reinstatement experiments reported in the same study: 0.5mA vs 3mA), and therefore may not have been perceived as aversive stimuli by the animals \cite{baldi2004footshock}. Without valid reminder shocks, the test trials would simply reflect the effect of extinction after a short delay (similar to the long-term memory test in the spontaneous recovery experiment in Gershman et al. \cite{gershman2013gradual}). Indeed, the gradual extinction group showed a higher conditioned response than the standard extinction group in the memory test in Gershman et al. \cite{gershman2013gradual} ($t(30)=3.05, p<.005$), consistent with the findings by Rescorla \cite{rescorla1975reinstatement}. In fact, simulating our model with low-intensity reinstatement trials showed minimal return of fear in subsequent tests across all three procedures (Figure \ref{fig:other_exp}C). Given the binary definition of shocks in our model, we modeled the low-intensity trials as no-shock trials, which could reflect how animals perceived such shocks due to the nonlinear psychophysics of pain perception \cite{le2001animal}.

Our model simulation suggests that an alternative way to reduce long-term fear memory is to gradually reduce the intensity of shocks (Figure \ref{fig:other_exp}D). Similar to the gradual extinction procedure used in \cite{gershman2013gradual}, the gradual reduction of intensity can facilitate the inference of one single latent cause. Specifically, the model predicts that the change in shock intensity needs to be gradual in the beginning of the extinction session (gray curve in Figure \ref{fig:other_exp}D); otherwise, animals will still show return of fear (yellow and purple curves). According to our model prediction, this new procedure is similarly effective as the original gradual-extinction procedure (green curve). However, a quantitative comparison depends on the values of model parameters. Empirical work is needed to test the new procedure and compare it with the original gradual extinction procedure.

\subsection{Limitations: differences between model predictions and empirical results}

Although the current model captures key aspects of the empirical findings, its predictions deviate from the experimental data quantitatively. First, the model correctly predicts the comparative differences between the three extinction conditions, but fails to predict the absolute return-of-fear effects. It over-predicts the reduction of fear for gradual extinction and gradual reverse, as compared to the empirical results (Figure \ref{fig:test_effect}). In fact, the model predicts lower fear responses in gradual extinction at test, and no change in gradual reverse when compared to the end of extinction, whereas animals showed minimal change and an increase in fear in these two conditions, respectively. Similarly, the predicted freezing rate during extinction deviates from the empirical results quantitatively (Supplementary Figure \ref{fig:overall_result_comparison}).

However, it is worth noting that behavioral variability was profound in these experiments. For instance, despite having the exact same procedures in conditioning and extinction sessions, the extinction effects differed in spontaneous recovery and reinstatement experiments (Supplementary Figure \ref{fig:overall_result_comparison}). Specifically, during the last four trials of extinction, freezing rate was significantly different between the two experiments under the same standard extinction procedure (one-way ANOVA: $F(1,22)=4.38, p<.05$; there were no significant difference for gradual extinction or gradual reverse procedures). These potentially different freezing rates at the end of extinction, nevertheless, served as the baselines for comparing test behavior in Figure \ref{fig:test_effect}, and provided the ``ground truth'' for comparison with our simulation results. Given this variability across animals and between experiments in the empirical findings, we decided to forego matching the empirical results quantitatively, and instead focused on correctly predicting the comparative difference between extinction procedures: gradual extinction being the most effective in reducing the return of fear, compared to either standard extinction or gradual reverse.

Quantitative deviations between the model's predictions and the empirical results may also be due to non-linear mapping between estimated shock probability and animals' freezing behavior, which we did not model. Because the form of this mapping was not the focus of the current work, we assumed a linear function for simplicity, but this is likely incorrect. Future work can design targeted experiments to investigate the underlying mechanism of this mapping.

Finally, the model captures behavior at the group level but does not make predictions regarding individual differences, which were marked in the empirical results (also observed in similar studies in humans \cite{gershman2015individual}). To make individual predictions with the current model, we can use different parameter values for each animal to capture their distinct learning, inference and behavioral processes. For example, animals often demonstrate abrupt switching (rather than gradual changes) in behavior that only seem gradual at the group level because different animals switch behavior at different times \cite{gallistel2004learning}. This between-animal variability in change points may be the result of different mapping functions from shock prediction to freezing rate, and may be an alternative to the perseveration assumption in explaining the gradual change of behavior over time. Another possibility is that individual animals do not perform full Bayesian inference but only take a small number of noisy samples from the probability distribution; in this case, aggregating over a group of animals will result in average behavior that resembles full inference, all the while each individual shows what seems like an idiosyncratic pattern of behavior \cite{daw2008pigeon}. Under this explanation, the current model is only adequate for explaining group-level behavior, and we need an additional sampling model to capture individual behavior. Both accounts for individual differences are likely in the task we modeled. We leave for future work to examine these sources of variability by fitting models to individual animals' behavior and comparing their predictions.

\subsection{Conclusion}

In sum, our work explains the effectiveness, or lack thereof, of different behavioral manipulations aimed at reducing maladaptive fear responses. Our results suggest that in acquiring and extinguishing fear responses, animals form a dynamic model of the environment, using inference of latent causes to predict future events. When representing and memorizing past experiences, however, our model suggests that animals summarize previous inference by collapsing distributions over potential latent-cause explanations and preserving the most likely explanation. Our findings suggest that, even in simple Pavlovian tasks such as fear extinction, animals' behavior can reveal a rich array of mechanisms of learning and representation.

\section*{Acknowledgements}
The authors would like to thank Lili X. Cai for assistance with the scoring neural network and helpful discussions, and Samuel J. Gershman for helpful comments on the draft.
The contents of this paper do not represent the views of the U.S. Department of Veterans Affairs or the United States Government. 

\section*{Data and Code Availability}
All data and code used in this paper are available at \url{https://github.com/mingyus/fear-extinction-latent-cause-inference}.

\printendnotes

\bibliography{references}

\newpage

\begin{suppfigure}[h!]
\centering
\includegraphics[width=\textwidth]{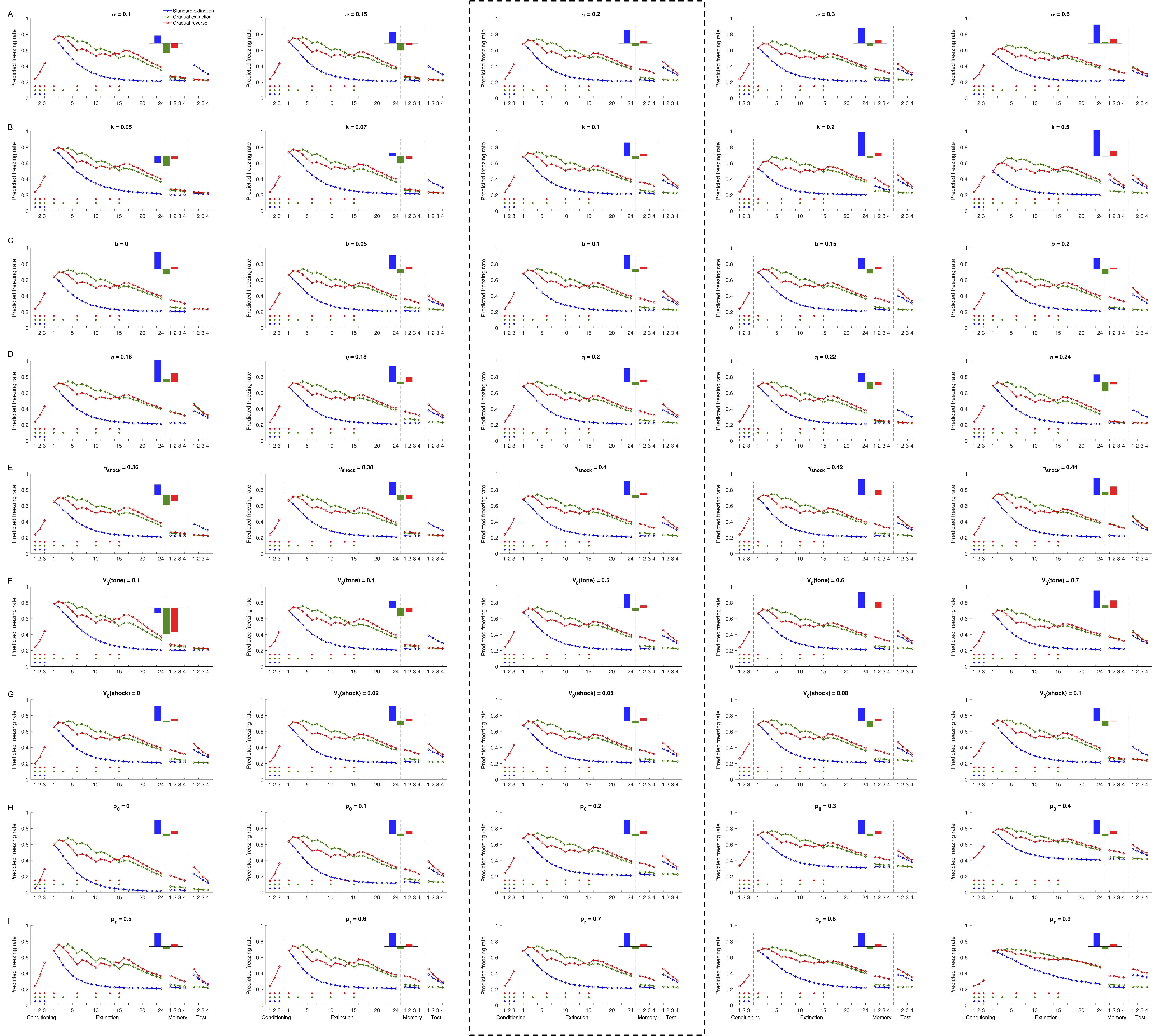}
\caption{\textbf{Effects of different parameter values on the predicted pattern of results.} We simulated the model by setting the values of all but one parameter as specified in Methods, and varying one parameter at a time. Each row of plots represents change of one parameter (as noted below and in sub-figure titles). For simplicity, we show only the spontaneous recovery experiment; results are consistent for the reinstatement experiment.
Within each sub-figure, the main graph shows the predicted freezing rate over the course of the experiment (similar to Figure \ref{fig:pshock_curve}A), and the inset bar graph shows the log posterior ratio between two-cause and one-cause assignments on the last trial of extinction (ranging from -4 to 4, the same scale across all sub-figures), with positive and negative values showing an advantage of the two-cause assignment and the one-cause assignment, respectively; extinction procedures color-coded as in the main graph. The center column (same for all rows), in a dashed square, corresponds to the original parameter values as in the main model: main graphs in this column are the same as Figure \ref{fig:pshock_curve}A; insets show the same results as in Figure \ref{fig:pcause}A, but only for the last extinction trial and in a bar graph.
\textbf{(A)} $\alpha = [0.1, 0.15, 0.2, 0.3, 0.5]$ (from left to right, same below). $\alpha$ is the concentration parameter for the distance-dependent Chinese restaurant process (ddCRP) prior, which controls the prior probability of generating a new latent cause. In the current experiments, $\alpha$ determines the \textit{a priori} probability ratio between forming two distinct causes and one single latent cause, shifting the curves in Figure \ref{fig:pcause}A up (larger $\alpha$'s) or down (smaller $\alpha$'s), and therefore changing the final posterior ratio (as shown in inset). With smaller $\alpha$'s (e.g., $\alpha = 0.1, 0.15$), animals form only one latent cause in the gradual reverse condition, and therefore show no return of fear in that condition; large $\alpha$ values (e.g., $\alpha = 0.5$) lead to the two-cause beliefs for all extinction procedures, and therefore similar return of fear across all conditions. \\
\textit{(continues on next page)}}
\label{fig:newp}
\end{suppfigure}

\begin{suppfigure}[h]
    \ContinuedFloat
    \caption[]{
    \textit{(continued from previous page)}\\[0.2em]
\textbf{(B)} $k = [0.05, 0.07, 0.1, 0.2, 0.5]$. $k$ is the decay parameter in the temporal kernel in the distance-dependent Chinese restaurant process prior, with a larger $k$ corresponding to a smaller reactivation probability for an old latent cause. As a result, smaller $k$'s (e.g., $k$ = 0.05) make one-cause assignment more likely across all procedures, and therefore result in no return of fear. As $k$ increases, two-cause beliefs first obtains dominance in standard extinction ($k = 0.07$), then in gradual reverse ($k = 0.1, 0.2, 0.5$), and eventually in gradual extinction if $k$ is large enough (not shown).
\textbf{(C)} $b = [0, 0.05, 0.1, 0.15, 0.2]$. b is the baseline probability for old causes in the distance-dependent Chinese restaurant process. Model simulation results are relatively insensitive to the value of b, unless very small (e.g., $b = 0$). In that case, all old causes become very unlikely after 30 days, so animals would infer a new latent cause during the spontaneous recovery test for all extinction procedures, resulting in no return of fear.
\textbf{(D,E)} $\eta = [0.16, 0.18, 0.2, 0.22, 0.24]$, and $\eta_\text{shock} = [0.36, 0.38, 0.4, 0.42, 0.44]$, respectively. $\eta$ and $\eta_\text{shock}$ are learning rates in Rescorla-Wagner update rule (the presence of shocks has a higher learning rate of $\eta_\text{shock}$, compared to all other stimuli, including ``tone'', ``no-tone'' and ``no-shock''). Learning rates determine the extent to which animals update their expectation after observing each stimulus. Specifically, in the current experiment, the comparison between $\eta_\text{shock}$ and $\eta$ is important in latent-cause inference. If animals learn about shocks with a relatively higher learning rate (corresponding to a higher value of $\eta_\text{shock}$ or a lower value of $\eta$), they will be more likely to infer two latent causes (each with distinct shock frequencies, thanks to faster update), resulting in return of fear for all procedures; to the contrary, a lower $\eta_\text{shock}$ or a higher $\eta$ lead to slower updates and animals tend to infer one latent cause (i.e., no return of fear in the gradual reverse condition).
\textbf{(F,G)} $V_0(\text{tone}) = [0.1, 0.4, 0.5, 0.6, 0.7]$, and $V_0(\text{shock}) = [0, 0.02, 0.05, 0.08, 0.1]$, respectively. $V_0(\text{tone})$ and $V_0(\text{shock})$ are the \textit{a priori} frequency of tones and shocks for a newly inferred latent cause. We had set the parameter values according to the relative frequency of the two types of stimuli in animals' natural environments (tones are presumably more common than shocks).
\textbf{(F)} If, however, the tone is assumed to be rarer for a new latent cause (i.e., lower values of $V_0(\text{tone})$), animals would be more likely to group all experiences into a single latent cause that is high in tone frequency. This is because all the trials are similar in having a tone, whereas a new cause would predict much lower tone frequency. As a result, the model predicts no return of fear for all procedures. To the contrary, if tones are assumed to be almost always present (e.g., $V_0(\text{tone}) \geq 0.7$), animals tend to infer new causes, which results in a two-cause belief and return of fear across all procedures. 
\textbf{(G)} Model simulation results are robust to small $V_0(\text{shock})$ values. If $V_0(\text{shock})$ is unrealistically high, however, animals would be unlikely to infer a second latent cause, as this new cause  will not be able to account well for the ``safe'' experience. The dominance of the one-cause belief would then lead to no return of fear in the gradual reverse condition.
\textbf{(H)} $p_0 = [0, 0.1, 0.2, 0.3, 0.4]$. $p_0$ is the baseline freezing rate after animals become aware of shocks. It does not affect latent-cause inference (i.e., the inset figures), but simply shifts the freezing rate curves up or down, in accordance with shifted baseline freezing seen empirically.
\textbf{(I)} $p_r = [0.5, 0.6, 0.7, 0.8, 0.9]$. $p_r$ is the perseveration parameter that controls the probability that animals will repeat the action from the previous trial. $p_r$ does not affect latent-cause inference, but rather determines the ``smoothness'' of the behavior curves.}
\end{suppfigure}

\begin{suppfigure}[ht!]
\centering
\includegraphics[width=0.8\textwidth]{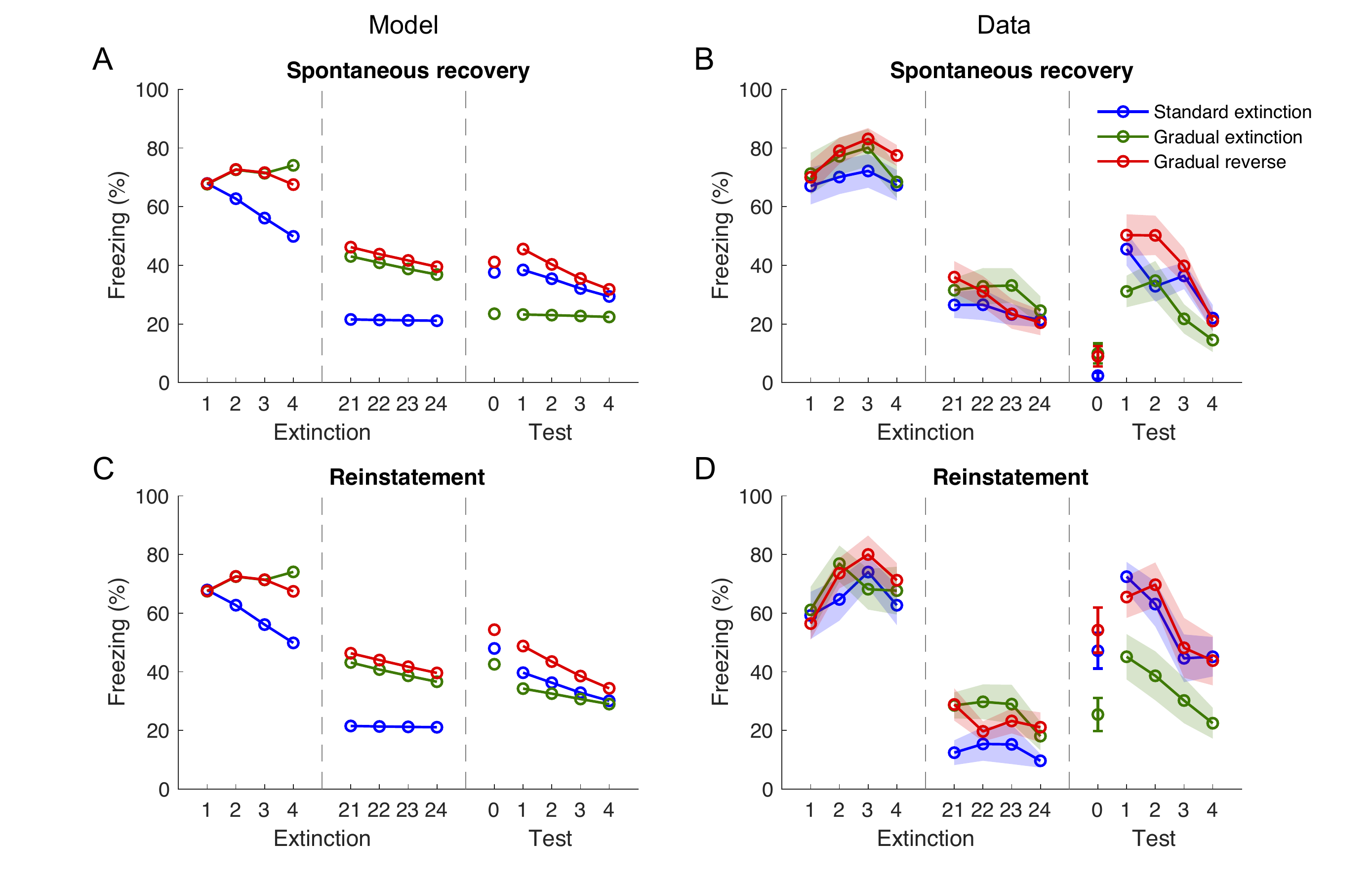}
\caption{\textbf{Freezing rate comparison between model predictions (A,C) and empirical results (B,D).} Shown are the first and last four trials of extinction, the beginning of the test session (before the first test trial; marked as trial 0), and the four test trials. Panels B and D are reproduced from \cite{gershman2013gradual}. Error bars and shaded areas in B,D indicate 1 s.e.m. across animals.}
\label{fig:overall_result_comparison}
\end{suppfigure}

\end{document}